\documentclass[%
 reprint,
superscriptaddress,
 amsmath,amssymb,
 aps,
floatfix,
]{revtex4-2}
\usepackage{graphicx}
\usepackage{dcolumn}
\usepackage{bm}
\usepackage[colorlinks,linkcolor=blue,anchorcolor=blue,urlcolor=blue, citecolor=blue]{hyperref}
\usepackage{float}
\usepackage{physics}
\usepackage{lipsum}  
\usepackage{mathtools}
\usepackage{array}
\usepackage{diagbox}
\usepackage{caption}
\usepackage{siunitx}
\usepackage{placeins}
\usepackage{afterpage}
\usepackage[table]{xcolor}
\usepackage{etoolbox}
\usepackage{multirow}
\usepackage{xr}
\usepackage{subfigure}
\usepackage{comment}
\usepackage{todonotes}

\newcommand{\DTUAffiliations}{
\affiliation{Center for Macroscopic Quantum States (bigQ), Department of Physics, Technical University of Denmark}
}

\newcommand{\DFMAffiliations}{
\affiliation{Danish Fundamental Metrology, Kogle Allé 5, 2970 Hørsholm, Denmark}
}

\newcommand{\prlsec}[1]{\textit{#1.}---\ignorespaces}

\begin{document}

\preprint{APS/123-QED}

\title{Long-distance continuous variable entanglement distribution over deployed fibre}

\author{Huy Q. Nguyen}%
\thanks{These authors contributed equally to this work.}
\DTUAffiliations
\author{Benjamin L. Larsen}%
\thanks{These authors contributed equally to this work.}
\DTUAffiliations
\author{Luiz C. C. P. Filho}
\DFMAffiliations
\author{Yijian Meng}
\DFMAffiliations
\author{Mikael Ø. Lassen}
\DFMAffiliations
\author{Jonas S. Neergaard-Nielsen}%
\DTUAffiliations
\author{Tobias Gehring}%
\DTUAffiliations
\affiliation{Celare Quantum Communications ApS, Frederiksberg, Denmark}
\author{Ulrik L. Andersen}
\DTUAffiliations
\author{Adnan A.E. Hajomer}
\DTUAffiliations

\date{\today}

\begin{abstract}
 Continuous-variable (CV) entanglement offers deterministic generation and room-temperature coherent detection for quantum networks, but distributing it over deployed fibre has been hindered by the need to stabilise optical phase, polarisation, and timing at remote receivers. Here we demonstrate, to our knowledge, the first distribution of CV entanglement over deployed metropolitan fibre using independent coherent receivers. The two modes of a two-mode squeezed vacuum state are delivered to remote nodes through 23 km
($\approx 7$\,dB loss) and 450 m ($\approx 1$\,dB loss) deployed fibre links. Each node uses its own free-running laser and polarisation-diverse coherent receiver. Co-propagating pilot tones enable digital recovery of timing, phase and polarisation, while inter-mode correlations fix the quadrature alignment. We certify entanglement with a Simon-Duan value of $1.915\pm 0.013$, below the separability bound of 2 and a logarithmic negativity of $0.067\pm0.011$. By replacing optical stabilisation with digital reconstruction, our approach offers a practical route towards scalable CV quantum networks.
\end{abstract}

\maketitle

\begin{figure*}[t]
    \centering
    \includegraphics[width=0.90\linewidth]{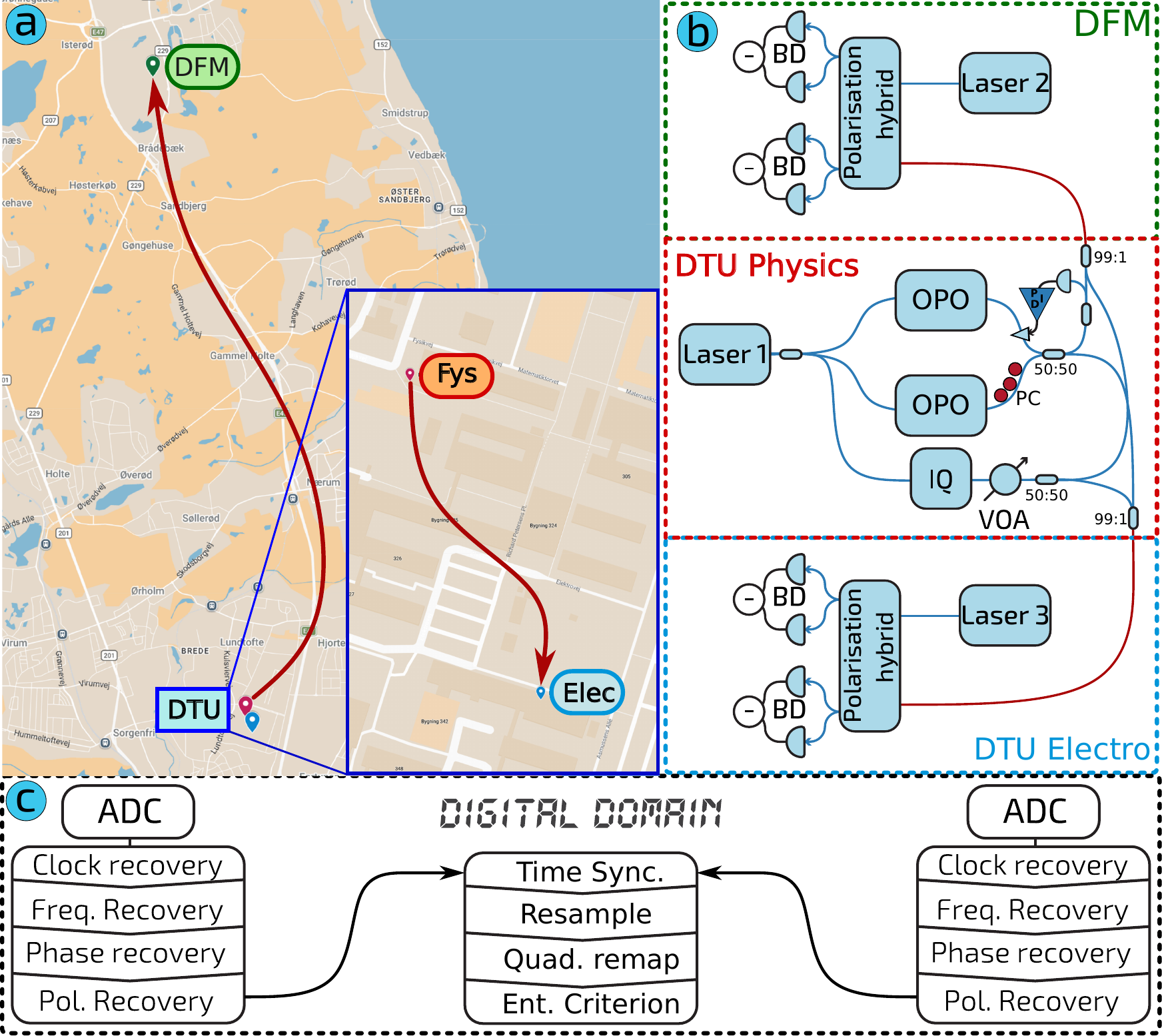}
    \caption{\textbf{Asynchronous distribution of CV entanglement across a metropolitan network.} The entanglement source at DTU Physics distributes the two modes of a TMSV state to the Danish National Metrology Institute (DFM) through 23 km of deployed fibre
($\approx 7$\,dB loss) and to DTU Electro through 450 m of campus fibre ($\approx 1$\,dB loss). Deployed links are shown in red. OPO: optical parametric oscillator; PC: polarisation controller; IQ: in-phase and quadrature modulator; VOA: variable optical attenuator; PID: proportional--integral--derivative controller; BD: balanced detector;
ADC: analogue-to-digital converter. Map data \copyright 2026 Google.}
    \label{fig:setup} 
\end{figure*}

Entanglement is a fundamental resource for quantum information technologies~\cite{horodeckiQuantum2009}, and its distribution between distant nodes is essential for building quantum networks~\cite{zhangEntanglementbased2024a}. Entangled photons have been distributed over hundreds of kilometres of deployed fibre~\cite{wengerowskyEntanglement2019,wengerowskyPassively2020,neumannContinuous2022}, using probabilistic photon-pair sources and single-photon detectors. Continuous-variable (CV) entanglement, such as two-mode squeezed vacuum (TMSV) states,  instead offers deterministic generation and high-bandwidth measurement using coherent receivers~\cite{weedbrookGaussian2012,lvovskySqueezed2015,kawasakiRealtime2025}, the technology underpinning modern optical communications. These features make CV entanglement particularly attractive for integration with existing telecommunications infrastructure and position it as a key resource for quantum teleportation~\cite{furusawaUnconditional1998}, measurement-based quantum computing~\cite{yoshikawaInvited2016,larsenDeterministic2021}, distributed quantum gates~\cite{filipContinuousvariable2004,yokoyamaNonlocal2014}, and quantum key distribution~\cite{madsenContinuous2012,gehringImplementation2015a}. 

Distributing CV entanglement, however, requires accessing quantum correlations between field quadratures measured at spatially separated nodes. For a TMSV state, each individual mode is locally thermal and the entanglement is encoded in correlations between the two modes. Accessing these correlations therefore requires establishing coherent phase reference between the remote coherent receivers. Conventionally, this is achieved by transmitting a strong local oscillator (LO) alongside the quantum signals or by optically phase-locking separate LO lasers~\cite{furusawaUnconditional1998,bowenExperimental2004,suleiman40KmFiber2022}. Both approaches become challenging in deployed fibre networks. Transmitting a strong LO consumes additional channel resources (temporal, spectral, or polarisation modes), risks leakage into the quantum signals, and can introduce security vulnerabilities in quantum cryptographic protocols~\cite{huangQuantum2013a,jouguetPreventing2013a}. Meanwhile, optical phase locking separate lasers requires additional hardware and active compensation of fibre-induced phase fluctuations, significantly complicating the system~\cite{suleiman40KmFiber2022}. Polarisation drifts during fibre transmission~\cite{nakamuraLongterm2024} and independent receiver clocks further complicate operation between remote nodes. Consequently, CV entanglement distribution has remained largely confined to laboratory fibre spools, relying on transmitted LOs and active optical stabilisation~\cite{huoDeterministic2018,yanGaussian2026}. Over deployed fibre, two-mode squeezing distribution was investigated across a short 1.5-km link using a transmitted LO~\cite{chapmanTwomode2023}, but entanglement between the remote nodes was not explicitly certified.

Coherent optical communication has overcome similar transmission impairments by shifting signal recovery from optical hardware to digital signal processing (DSP), which compensates for frequency offsets, phase noise, polarisation transformations, and timing errors after detection. This approach has driven major advances in high-speed and long-distance CV quantum key distribution (CV-QKD)~\cite{hajomerChipbased2026,hajomerCoexistence2025,wangHighrate2025} and has recently been extended to the digital reconstruction of squeezed light and practical squeezed state CV-QKD~\cite{nguyenDigital2025a,nguyenPractical2026}. Extending these techniques to distribute entanglement, however, presents a fundamentally different challenge. Unlike single-mode squeezing, where the noise reduction appears in the quadrature measurement of a single optical field, each TMSV mode is locally thermal, so the relative quadrature alignment between independent receivers cannot be determined from local measurements alone.

Here we demonstrate CV entanglement distribution over a deployed metropolitan fibre network using independent coherent receivers. Pilot-assisted digital recovery of timing, frequency, phase and polarisation, combined with correlation-based quadrature alignment, removes the need for receiver-side optical phase locking, active polarisation control and shared sampling clocks. In the laboratory, phase recovery reveals entanglement through logarithmic negativity, whereas the Duan–Simon criterion is violated only after quadrature alignment. Over deployed links of 23 km and 450 m, we certify entanglement between remote nodes, to our knowledge the first deployed-fibre CV entanglement distribution. By moving the recovery of the relative measurement frame from optical stabilisation to digital processing, our work establishes a practical architecture for distributing CV entanglement over existing telecommunications infrastructure.

\begin{figure*}[t]
    \centering
    \includegraphics[width=0.90\linewidth]{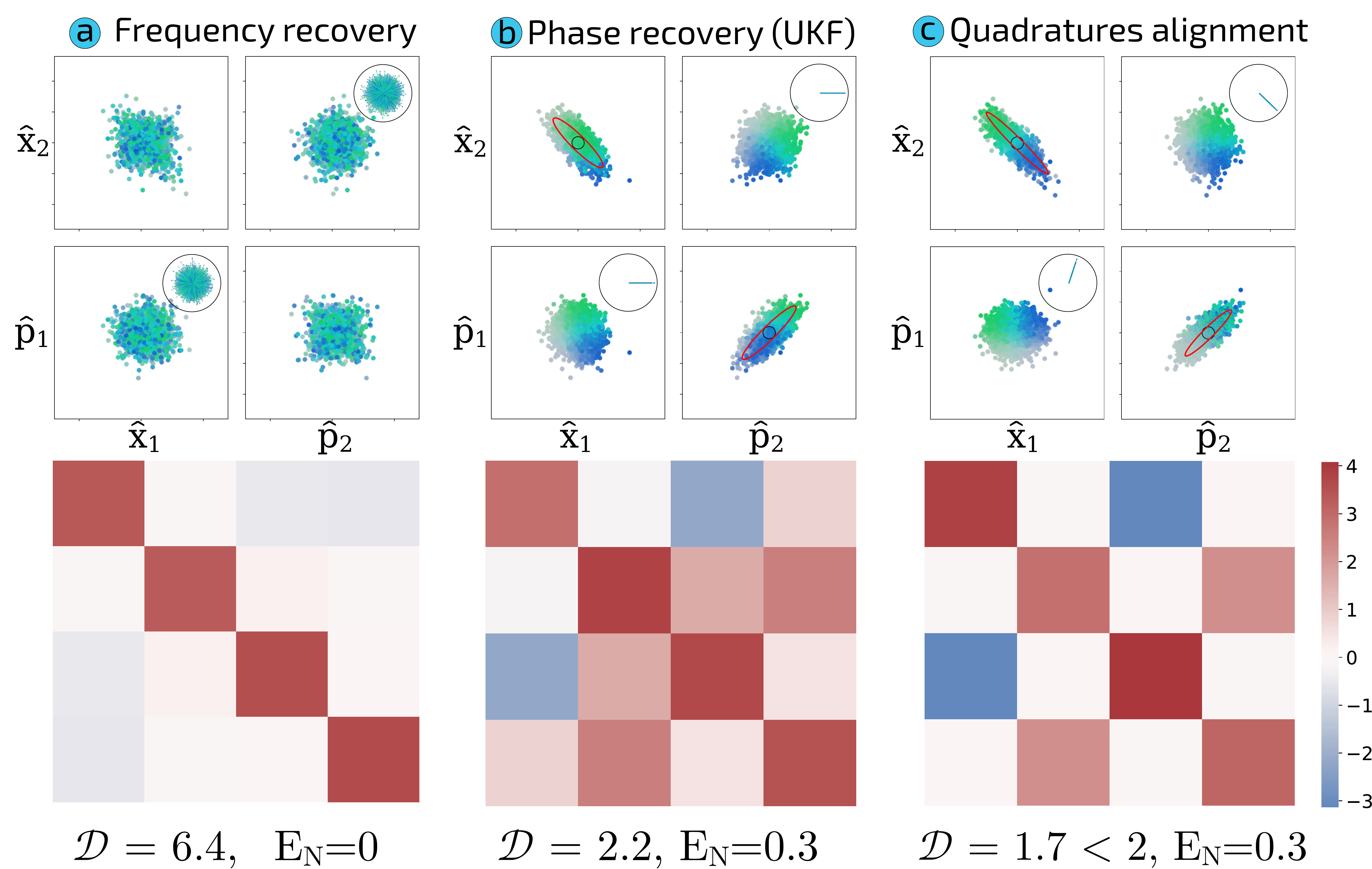}
    \caption{\textbf{Laboratory demonstration.} The setup is described in Supplementary Material~\ref{supp:labsetup} (Fig.~\ref{fig:catlabsetup}). The TMSV is measured using an independent laser as the LLO. Top: phase-space representations of the entangled state, each showing 2500 measurements (the circular insets show the phase of the pilot tone). Bottom: the corresponding covariance matrices, each computed from a single measurement sequence of $\approx5\times10^4$ measurements$\mathcal{D}$:~Simon-Duan value, $E_N$: entanglement log-negativity. 
    }
    \label{fig:DSPExample}
\end{figure*}

\prlsec{Method}
We generate TMSV states at 1550\,nm by interfering two single-mode squeezed vacuum states (SMSVs) on a balanced beam splitter. The squeezed states are produced by hemilithic optical parametric oscillators (OPOs), each containing a type-0 periodically poled KTP crystal pumped at 775\,nm, with a linewidth (FWHM) of approximately 60\,MHz. The OPO cavities are locked using the H\"ansch--Couillaud technique~\cite{hanschLaser1980}, and the relative phase of the squeezed states is locked using a 1\% tap after the beam splitter. An alternating lock-and-measure scheme, implemented with acousto-optic modulators, blocks the coherent probe beams during measurement.

Two classical pilot tones at $\pm100$\,MHz around the optical carrier are generated with an in-phase and quadrature (IQ) modulator, combined with the two entangled modes through 99:1 beam splitters. Since the pilots lie outside the OPO bandwidth, they contribute no in-band excess
noise. The modes are transmitted over deployed fibre to two remote stations, each using an independent, free-running laser as its locally generated LO (LLO), detuned by more than 100 MHz from the signal. Arbitrary polarisation transformations in the fibre are handled without active optical control by polarisation-diverse coherent receivers, in which each polarisation component is measured by a balanced photodetector. Radio-frequency (RF) heterodyne detection gives access to both quadratures of each mode~\cite{nguyenDigital2025a}. The detector outputs are digitised by independent 16-bit analogue-to-digital converters at 1\,GSa/s with no shared sampling clock. The vacuum noise level is calibrated by one-shot-noise calibration~\cite{ZhangOneShot2020}, with $> 10$ dB electronic-noise clearance.

\begin{figure*}[t]
  \centering
    \includegraphics[width=0.9\linewidth]{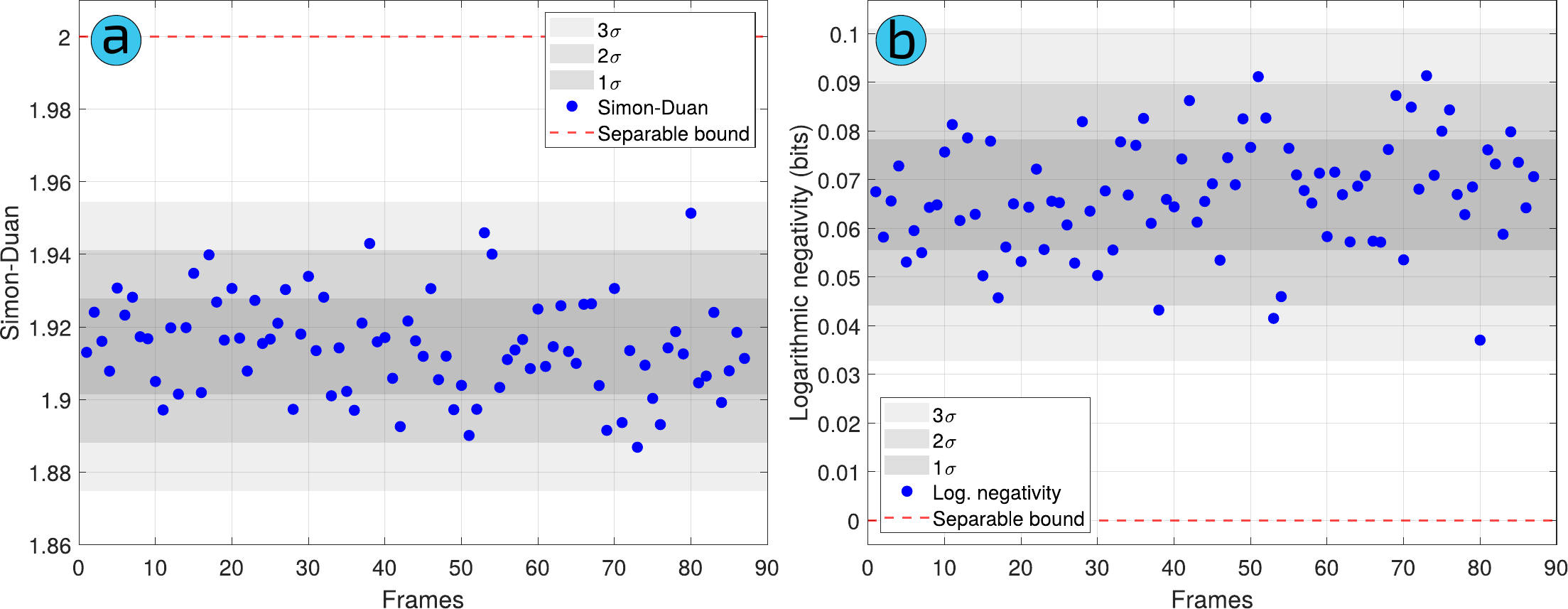}
\caption{\textbf{Field demonstration.} (a)~Simon--Duan value $\mathcal{D}$ and (b)~logarithmic negativity $E_N$ for 87 consecutive frames of $4.5\times10^4$ measurements each. Shaded bands:1$\sigma$--3$\sigma$ frame-to-frame spread; dashed lines: separability
bounds.}
  \label{fig:entanglement}
\end{figure*}

The DSP chain, shown in Fig.~\ref{fig:setup}(c), reconstructs the measurement records acquired independently at the two stations. First, the sampling-clock offset is estimated from the measured separation of the two pilot tones, which is nominally 200\,MHz at the transmitter and insensitive to the LO frequency offset~\cite{nguyenDigital2025a,chinDigital2022}. The frequency and phase offsets between each LO and the quantum signal are then estimated separately for each polarisation component using an unscented Kalman filter (UKF)~\cite{chinMachine2021}. Removing the relative phase between the two polarisation components reduces the elliptical polarisation transformation introduced by the fibre to a linear one, which is then undone by a digital rotation. Finally, the signals are filtered, aligned in time using their cross-correlation and resampled to 10\,MSa/s to obtain the quadrature measurements.

Because each mode of a TMSV state is locally thermal, its quadrature orientation cannot be inferred from local measurements alone. We therefore recover the relative quadrature alignment from the inter-mode correlations. For an aligned TMSV state, the zero-mean cross-quadrature correlations vanish, $\langle\hat{x}_1\hat{p}_2\rangle=\langle\hat{p}_1\hat{x}_2\rangle=0$.
We determine the local rotation angles by minimising 
\begin{equation}
    (\theta^*,\phi^*) = \arg\min_{\theta,\phi} \Big( \big|\langle \hat{x}_1'\,\hat{p}_2' \rangle\big| + \big|\langle \hat{p}_1'\,\hat{x}_2' \rangle\big| \Big)^2 ,
    \label{eq:cost_angle}
\end{equation}
where the rotated quadratures are:
\begin{equation}
\begin{aligned}
    \hat{x}_1' &= \hat{x}_1\cos\theta - \hat{p}_1\sin\theta, &
    \hat{p}_1' &= \hat{x}_1\sin\theta + \hat{p}_1\cos\theta, \\
    \hat{x}_2' &= \hat{x}_2\cos\phi - \hat{p}_2\sin\phi, &
    \hat{p}_2' &= \hat{x}_2\sin\phi + \hat{p}_2\cos\phi.
\end{aligned}
\end{equation}

Here, $\theta$ and $\phi$ are the rotations applied to modes 1 and 2, respectively. For a perfectly symmetric TMSV, the cost function depends only on the sum $\theta+\phi$, so one can fix $\theta = 0$ and optimise over $\phi$ alone. In practice, however, the TMSV can be asymmetric due to different anti-squeezing levels between the two SMSVs, and $\theta$ and $\phi$ must therefore be optimised simultaneously. One caveat is that the optimisation can converge to a configuration where one mode is rotated by $\pi/2$, i.e., $\hat{x}_1$ becomes aligned with $\hat{p}_2$; this is easily corrected if needed. For an asymmetric TMSV, the ambiguity can be resolved by adopting a fixed convention, for example, always aligning the modes so that the $\hat{x}$-quadrature variance is lower than that of $\hat{p}$.

We characterise the reconstructed entanglement using the
Simon--Duan inseparability criterion~\cite{duanInseparabilityCriterionContinuous2000,simonPeresHorodecki2000}
and the logarithmic negativity~\cite{vidal2002computable}.
For separable states, the normalised Simon--Duan quantity satisfies
\begin{equation}
\mathcal{D} =
\min_{a\neq 0}
\frac{\left[
\langle(\Delta\hat{u})^2\rangle+
\langle(\Delta\hat{v})^2\rangle
\right]}{a^2+a^{-2}}
\geq 2,
\label{eq:duan}
\end{equation}
where $\hat{u}=|a|\hat{x}_1+a^{-1}\hat{x}_2$ and
$\hat{v}=|a|\hat{p}_1-a^{-1}\hat{p}_2$, consistent with the
anti-correlated $\hat{x}$ and correlated $\hat{p}$ quadratures
fixed by our alignment convention. The logarithmic negativity is
$E_N=\max(0,-\log_2\tilde{\nu}_-)$, where $\tilde{\nu}_-$ is the
smallest symplectic eigenvalue of the partially transposed
covariance matrix~\cite{vidal2002computable}. Unlike the
Simon--Duan quantity, which is evaluated in a fixed quadrature
basis, the logarithmic negativity is invariant under local
phase-space rotations.

\prlsec{Laboratory demonstration}
We first investigate the digital reconstruction of entanglement in the laboratory, without deployed fibre, using the setup described in Supplementary Sec.~\ref{supp:labsetup} (Fig.~\ref{fig:catlabsetup}). Since both modes are digitised by a common ADC, and the polarisation is aligned with a polarisation controller that remains stable over the short laboratory fibres, clock synchronisation and polarisation recovery are not required. Carrier frequency and phase recovery is therefore applied once, to a single polarisation component, and is followed by quadrature alignment as described above. Figure~\ref{fig:DSPExample} shows the measured phase-space distributions and the corresponding covariance matrices after successive stages of digital processing, with circular insets showing the pilot-tone phase extracted at each stage. Each point is coloured by the phase-space quadrant it occupies after full reconstruction, i.e., where it would be measured at the source in the absence of phase noise, and keeps this colour at every processing stage. After frequency recovery (Fig.~\ref{fig:DSPExample}(a)), random phase fluctuations between the quantum signal and the local oscillator (inset) wash out the inter-mode correlations, yielding $\mathcal{D}=6.4$ and $E_N=0$. Digital phase recovery establishes a stable phase reference across all measured states (Fig.~\ref{fig:DSPExample}(b), inset) and restores the inter-mode correlations, which now appear in the off-diagonal blocks of the covariance matrix. The logarithmic negativity already reaches $E_N=0.3$, yet the Simon--Duan quantity remains above the separability bound ($\mathcal{D}=2.2$). Part of the correlations still resides in the cross-quadrature terms $\langle\hat{x}_1\hat{p}_2\rangle$ and $\langle\hat{p}_1\hat{x}_2\rangle$, which the fixed-basis combinations in Eq.~(\ref{eq:duan}) do not capture. The correlation-based quadrature alignment of Eq.~(\ref{eq:cost_angle}) rotates these correlations into the $\{\hat{x}_1,\hat{x}_2\}$ and $\{\hat{p}_1,\hat{p}_2\}$ blocks (Fig.~\ref{fig:DSPExample}(c)), yielding $\mathcal{D}=1.7<2$, while $E_N$ remains unchanged. Since $E_N$ is invariant under local phase-space rotations, the entanglement is already fully present after phase recovery. The final alignment does not recover entanglement; it rotates the local measurement frames into the basis in which a fixed-quadrature witness, and protocols relying on specific quadrature correlations such as teleportation and QKD, can access it. 

\prlsec{Field demonstration}We then demonstrate entanglement distribution over a deployed metropolitan fibre network (Fig.~\ref{fig:setup}). The two modes of the TMSV state, generated at DTU Physics, are distributed to the Danish National Metrology Institute (DFM) through 23\,km of deployed fibre ($\approx 7$\,dB loss; mode~1) and to DTU Electro through 450\,m of campus fibre ($\approx 1$\,dB loss; mode~2). Both stations operate with independent free-running lasers and sampling clocks. After digital clock, frequency, phase and polarisation recovery, followed by quadrature alignment, we record the entanglement in 87 frames, each corresponding to one measurement period of the lock-and-measure cycle and containing $4.5\times10^4$ quadrature measurements (Fig.~\ref{fig:entanglement}). Every frame violates the separability bound, giving on average $\mathcal{D}=1.915\pm0.013$ and $E_N=0.067\pm0.011$ (one standard deviation across frames).

The covariance matrix obtained from approximately $3.9\times10^6$
measurements is
\begin{equation}
\gamma_{\mathrm{field}}=
\begin{bmatrix}
1.104 & 0.009 & -0.268 & 0\\
0.009 & 1.074 & 0 & 0.232\\
-0.268 & 0 & 1.448 & 0.042\\
0 & 0.232 & 0.042 & 1.392
\end{bmatrix},
\label{eq:final_covar}
\end{equation}
in shot-noise units with quadrature ordering $(\hat{x}_1,\hat{p}_1,\hat{x}_2,\hat{p}_2)$. The reconstructed covariance matrix exhibits the characteristic TMSV structure:
anti-correlated $\hat{x}$ and correlated $\hat{p}$ quadratures between the two nodes, with the cross-quadrature terms
$\langle\hat{x}_1\hat{p}_2\rangle$ and $\langle\hat{p}_1\hat{x}_2\rangle$ set to zero by the alignment of
Eq.~(\ref{eq:cost_angle}). The local variances of mode~1 are lower than those of mode~2, reflecting the higher loss of the 23 km link.

\prlsec{Conclusion}
We have demonstrated CV entanglement distribution between independent remote nodes through 23 km and 450 m deployed fibre links, achieving, to our knowledge, the first deployed-fibre distribution of CV entanglement to date. Using independent coherent receivers with pilot-assisted digital recovery of timing, frequency, phase and polarisation, combined with correlation-based quadrature alignment, we certify entanglement with a Simon--Duan value of $1.915\pm0.013$ and a logarithmic negativity of $0.067\pm0.011$, despite the intrinsic 3\,dB loss associated with heterodyne detection. Our results show that the relative measurement frame required to reveal CV entanglement can be reconstructed digitally, without receiver-side optical phase locking, active polarisation control or shared sampling clocks. By enabling each receiver to operate with a local laser and independent sampling clock, this architecture offers a promising route towards multi-node CV quantum networks and potential integration with wavelength-multiplexed classical communications~\cite{hajomerCoexistence2025}. This work provides a foundation for extending entanglement-based quantum communication, teleportation and distributed sensing to independently operated nodes over existing telecommunications infrastructure.

\section*{Acknowledgments}
We thank GlobalConnect for installing the fibre link between DTU and DFM. This work was supported by the Danish National Research Foundation through the Center for Macroscopic Quantum States (bigQ, DNRF142); the European Union's Horizon Europe research and innovation programme through the Quantum Security Networks Partnership (QSNP, Grant Agreement No.~101114043); Innovation Fund Denmark through the CyberQ project (Grant Agreement No.~3200-00035B) and the AccessQKD project (Grant Agreement No.~43560-00004B); and the Carlsberg Foundation (Grant No.~CF21-0466).

\bibliography{apssamp}.

\clearpage
\newpage

\onecolumngrid
\section*{Supplementary information}
\setcounter{subsection}{0}
\renewcommand{\thesubsection}{S\arabic{subsection}}
\setcounter{figure}{0}
\renewcommand{\thefigure}{S\arabic{figure}}

\subsection{Expected loss}

For a pure two-mode squeezed vacuum state which undergoes loss of $\eta_A$ in one mode and $\eta_B$ in the other mode and a squeezing parameter $r$, the value of the Simon-Duan bound is. 

\begin{align*}
    \mathcal{D}_a = 2 \big( a^2 \eta_A \cosh(2r) + \frac{1}{a^2} \eta_B \cosh(2r) -2  \sqrt{\eta_A\eta_B} \sinh(2r) + a^2 (1-\eta_A) + \frac{1}{a^2} (1-\eta_B)\big)
\end{align*}

here we have assumed without loss of generality that the state is $\begin{bmatrix}
    \cosh(2r) \mathbb{I} & -\sinh(2r) \sigma_z \\
    -\sinh(2r) \sigma_z & \cosh(2r) \mathbb{I}
\end{bmatrix}$, thus $a>0$. 

For this model the optimal value of parameter $a$ can be found to be: 
\begin{align*}
    a_{opt}^2 = \sqrt{\frac{\eta_A\cosh(2r)+(1-\eta_A)}{\eta_B\cosh(2r)+(1-\eta_B)}}
\end{align*}

The efficiency of the different parts of the setup can be seen in Tab.~\ref{tab:lossBudget}. The total efficiency is thus: $\eta_A \approx 5\%$ and $\eta_B \approx 22 \%$. The cavities are pumped at 45\% of the threshold which gives $r=0.59$. For these parameters the expected value of the Simon-Duan is $\mathcal{D}_{a_{opt}} = 1.90$.

\begin{table}[ht]
    \centering
    \begin{tabular}{|c|c|c|c|c|c|c|}
        \hline 
        Escape efficiency &  Coupling into fibre & EPR generation & Fibre link A & Fibre link B & Intrinsic het. loss & Detector loss \\ \hline
        $\approx 98$ \% & $\approx 94$ \% & $\approx 98$ \% & $\approx $ -7 dB & $\approx$ -0.8 dB & -3 dB & $\approx 59 \%$ \\ \hline
    \end{tabular}
    \caption{Efficiencies of different steps, the escape efficiency, efficiency of coupling into fibre are quite similar for the two SMSV sources.}
    \label{tab:lossBudget}
\end{table}

\newpage

\subsection{Laboratory demonstration setup}\label{supp:labsetup}
\begin{figure*}[h!]
    \centering
    \includegraphics[width=0.65\linewidth]{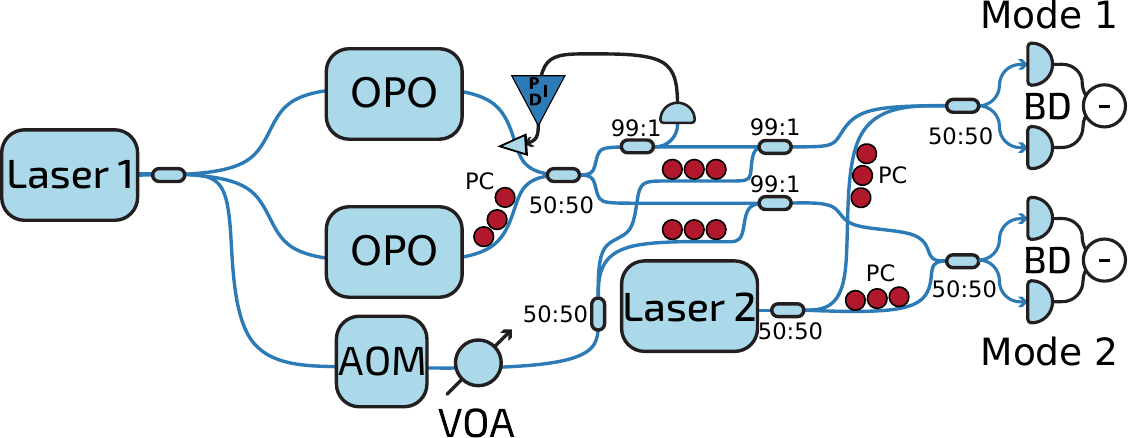}
\caption{\textbf{Laboratory demonstration setup.} The entanglement source is identical to that in Fig.~\ref{fig:setup}. A pilot tone generated by an acousto-optic modulator (AOM) at a 40-MHz offset is combined with each TMSV mode, and both quadratures of each mode are measured by RF-heterodyne detection using an LLO from an independent free-running laser.OPO: optical parametric oscillator; AOM: acousto optics modulator; VOA: variable optical attenuator; PC: polarisation controller; BD: balanced detector}
    \label{fig:catlabsetup} 
\end{figure*}

Figure~\ref{fig:catlabsetup} shows the setup used for the laboratory demonstration in the main text, with results shown in Fig.~\ref{fig:DSPExample}. The entanglement source is identical to that of the deployed experiment. Because both modes are digitised by a common ADC, clock recovery is not required and a single pilot tone suffices. We therefore generate the pilot with an acousto-optic modulator (AOM) at a 40-MHz frequency offset instead of with the IQ modulator, and match its polarisation to that of the TMSV using two polarisation controllers. Since the polarisation in the laboratory is stable, two further polarisation controllers match the polarisation of the LLO, generated by an independent free-running laser, to that of each TMSV mode. Both quadratures of each mode are then measured by RF-heterodyne detection and balanced detector.

\newpage
\subsection{Entanglement recovery in deployed fibre}

\begin{figure*}[ht]
    \centering
    \includegraphics[width=\linewidth]{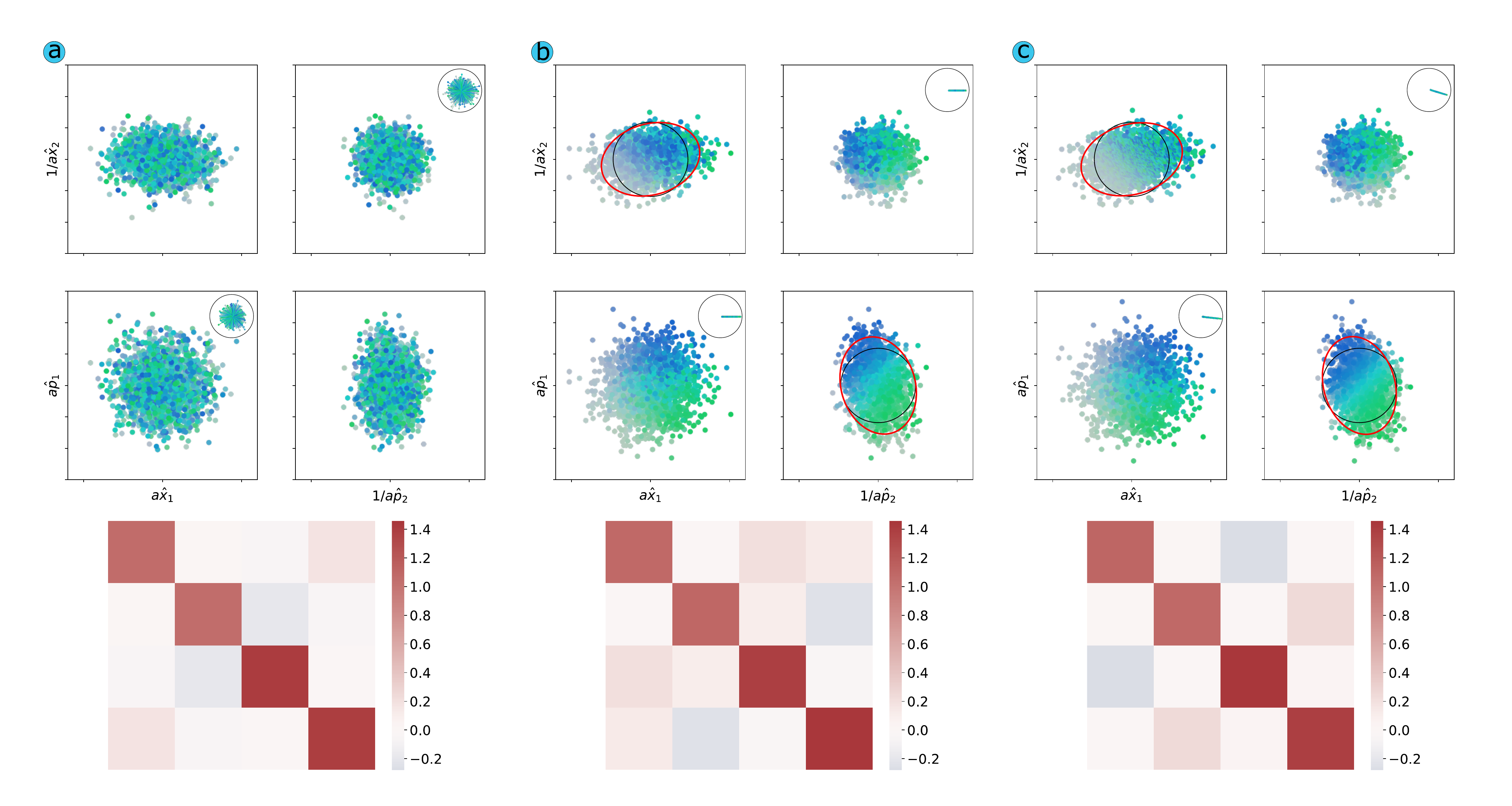}
    \caption{\textbf{Entanglement recovery in deployed fibre.} 
    Top: phase space representation, each showing 2250 measurements, notice that the axes have been scaled with the parameter $a$ which optimised violation of $\mathcal{D}$. Bottom, covariance matrices. 
    }
    \label{fig:DSPExampleDeployed}
\end{figure*}

Similar to Fig. ~\ref{fig:DSPExample} we show the phase space representation and covariance matrix of three steps in the entanglement recovery protocol in Fig. ~\ref{fig:DSPExampleDeployed}. Due to the asymmetric efficiencies in the deployed fibre we scale the axes in the phase space representation to be able to visualise the violation of the SD-bound. 

\end{document}